\documentclass[%
 reprint,superscriptaddress,
 amsmath,amssymb,
 aps, prapplied,
floatfix,
]{revtex4-2}

\usepackage{epsfig}
\usepackage[T1]{fontenc}
\usepackage{amssymb}
\usepackage{amsmath}
\usepackage{color}
\usepackage{upgreek}

\begin{document}


\title{Effective uniaxial dielectric function tensor and optical phonons in ($\bar{2}01$)-plane oriented $\beta$-Ga$_2$O$_3$ films with equally-distributed six-fold rotation domains}

\author{Alyssa Mock}
\affiliation{Department of Electrical and Computer Engineering, College of Engineering, Applied Science and Technology,Weber State University, Ogden, 84408, Utah, USA}
\author{Steffen Richter}
\affiliation{Solid State Physics and NanoLund, Lund University, Lund, 22100, Sweden}
\affiliation{Terahertz Materials Analysis Center (THeMAC) and Competence Center for III-Nitride Technology, C3NiT - Janz\'en,Link\"{o}ping University, Link\"{o}ping, 58183, Sweden}
\author{Alexis Papamichail}
\affiliation{Terahertz Materials Analysis Center (THeMAC) and Competence Center for III-Nitride Technology, C3NiT - Janz\'en,Link\"{o}ping University, Link\"{o}ping, 58183, Sweden}
\author{Vallery Stanishev}
\affiliation{Terahertz Materials Analysis Center (THeMAC) and Competence Center for III-Nitride Technology, C3NiT - Janz\'en,Link\"{o}ping University, Link\"{o}ping, 58183, Sweden}
\affiliation{Department of Physics, Chemistry and Biology (IFM), Link\"{o}ping University, Link\"{o}ping, 58183, Sweden}
\author{Misagh Ghezellou}
\affiliation{Department of Physics, Chemistry and Biology (IFM), Link\"{o}ping University, Link\"{o}ping, 58183, Sweden}
\author{Jawad Ul-Hassan}
\affiliation{Department of Physics, Chemistry and Biology (IFM), Link\"{o}ping University, Link\"{o}ping, 58183, Sweden}
\author{Andreas Popp}
\affiliation{Leibniz-Institut f\"ur Kristallz\"uchtung, 12489, Berlin, Germany}
\author{Saud Bin Anooz}
\affiliation{Leibniz-Institut f\"ur Kristallz\"uchtung, 12489, Berlin, Germany}
\author{Daniella Gogova}
\affiliation{Department of Physics, Chemistry and Biology (IFM), Link\"{o}ping University, Link\"{o}ping, 58183, Sweden}
\author{Praneeth Ranga}
\affiliation{Department of Electrical and Computer Engineering, The University of Utah, Salt Lake City, 84112, Utah, USA}
\author{Sriram Krishnamoorthy}
\affiliation{Department of Electrical and Computer Engineering, The University of Utah, Salt Lake City, 84112, Utah, USA}
\affiliation{Materials Department, University of California Santa Barbara, Santa Barbara, 93106-5050, California, USA}
\author{Rafal Korlacki}
\affiliation{Department of Electrical and Computer Engineering, University of Nebraska-Lincoln, Lincoln, 68588, Nebraska, USA}
\author{Mathias Schubert}
\affiliation{Department of Electrical and Computer Engineering, University of Nebraska-Lincoln, Lincoln, 68588, Nebraska, USA}
\affiliation{Solid State Physics and NanoLund, Lund University, Lund, 22100, Sweden}
\author{Vanya Darakchieva}
\affiliation{Solid State Physics and NanoLund, Lund University, Lund, 22100, Sweden}
\affiliation{Terahertz Materials Analysis Center (THeMAC) and Competence Center for III-Nitride Technology, C3NiT - Janz\'en,Link\"{o}ping University, Link\"{o}ping, 58183, Sweden}
\affiliation{Department of Physics, Chemistry and Biology (IFM), Link\"{o}ping University, Link\"{o}ping, 58183, Sweden}

\begin{abstract}
	Monoclinic $\beta$-Ga$_2$O$_3$ films grown on $c$-plane sapphire have been shown to exhibit six $(\bar{2}01)$-plane oriented domains, which are equally-spaced-by-rotation around the surface normal and equally-sized-by-volume that render the film optical response effectively uniaxial. We derive and discuss an optical model suitable for ellipsometry data analysis of such films. We model mid- and far-infrared ellipsometry data from undoped and electrically insulating films with an effective uniaxial dielectric tensor based on projections of all phonon modes within the rotation domains parallel and perpendicular to the sample normal, i.e., to the reciprocal lattice vector $\mathbf{g}_{\bar{2}01}$. Two effective response functions are described by model, and found sufficient to calculate ellipsometry data that best-match measured ellipsometry data from a representative film. We propose to render either effective dielectric functions, or inverse effective dielectric functions, each separately for electric field directions parallel and perpendicular to $\mathbf{g}_{\bar{2}01}$, by sums of Lorentz oscillators, which permit to determine either sets of transverse optical phonon mode parameters, or sets of longitudinal optical phonon mode parameters, respectively. Transverse optical modes common to both dielectric functions can be traced back to single crystal modes with $B_{\mathrm{u}}$ character, while modes with $A_{\mathrm{u}}$ character only appear within the dielectric function for polarization perpendicular to the sample surface. The thereby obtained parameter sets reveal all phonon modes anticipated from averaging over the six-fold rotation domains of single crystal $\beta$-Ga$_2$O$_3$, but with slightly shifted transverse optical, and completely different longitudinal optical phonon modes. Structural analysis of the film revealed virtually strain-free material. We suggest small crystal grains and high density of grain boundaries as a possible origin for the observed transverse optical phonon frequency shifts with respect to bulk material. The differences in longitudinal optical modes here compared to the bulk are hypothesized to be caused by averaging of the electric field induced polarization over many long-range ordered rotation domains. Our model can be useful for future analysis of free charge carrier properties using infrared ellipsometry on multiple domain films of monoclinic structure $\beta$-Ga$_2$O$_3$.
\end{abstract}

\maketitle



\section{Introduction}

Ultra-wide band gap conductive and semiconducting metal oxides have attracted substantial interest in high power electronics and short wavelengths optoelectronics applications \cite{Spencer2022}. 
Among the semiconducting oxides, gallium oxide stands out due to its large bandgap ($\approx$5\,eV) and high breakdown field (up to 8\,MV/cm) \cite{Pearton2018a,Pearton2018,Green2022}. 
Five known phases ($\upalpha$, $\beta$, $\upgamma$, $\updelta$, and $\upepsilon$ or $\upkappa$) exist, of which the thermodynamically stable, monoclinic $\beta$-phase has taken a focal point in ongoing semiconductor research. 
The low crystal symmetry is a cause for numerous new physical properties. 
For example, $\beta$-Ga$_2$O$_3$ is characterized by three lowest-energy direct optical transitions at the Brillouin zone center where each transition has a different energy, each polarized in different directions. Likewise, each transition dipole is polarized differently within the monoclinic unit cell \cite{Ratnaparkhe2017,Sturm2016,Mock2017}. 
Consequently, $\beta$-Ga$_2$O$_3$ is pleochroic, reveals different absorption behavior for different directions within the crystal and dispersion of optic axes which all have been studied in wide spectral ranges \cite{Mock2017,Sturm2015,Schubert2016}. 

$\beta$-Ga$_2$O$_3$ reveals complex interaction between the anisotropic lattice response and charge carriers \cite{Fiedler2020,Onuma2016,Ghosh2017,Schubert2019,Onuma2021}. 
In order to investigate these, thorough understanding of the phonon modes in the material is required. 
There are eight infrared-active optical phonon modes polarized within the monoclinic $ac$-plane ($B_{\mathrm{u}}$ modes) and an additional four which are polarized along the $b$-axis ($A_{\mathrm{u}}$ modes) \cite{Schubert2016,Villora2002}. 
The optical response for electric-field polarization within the monoclinic plane is complex and can be modeled by using an \textit{eigendielectric displacement} approach \cite{Sturm2016,Schubert2016,Schubert2016a,Schubert2019a}. 
Each optical phonon mode is associated with a pair of transverse optical (TO) and longitudinal optical (LO) resonances. 
Most notably, none of the TO and LO resonances within the monoclinic plane share a common polarization direction. Furthermore, the so-called TO-LO order rule was found violated in $\beta$-Ga$_2$O$_3$ for the modes polarized within the monoclinic plane \cite{Schubert2019a}. 
This is in contrast to high symmetry materials such as GaAs or GaN, where TO and LO modes always share common directions. Along such high symmetry directions the frequency order of TO and LO modes was always observed with first a TO mode at lower frequency followed directly by an LO mode at higher frequency. \cite{Schubert2019a,Schubert-book}. The unusual phonon order within the monoclinic plane of $\beta$-Ga$_2$O$_3$ was explained by the non-collinear appearance of phonons in low-symmetry materials. The subsequent occurrence of inner and outer nested modes causes the violation of the TO-LO rule \cite{Schubert2019a}. 

For doped $\beta$-Ga$_2$O$_3$, it was revealed that coupling of free charge carriers with the LO modes results in coupled phonon plasmon modes with a polarization direction that shifts with carrier concentration throughout the entire monoclinic plane \cite{Schubert2019}. 
Hence, charge carrier as well as thermal transport is expected to be highly anisotropic as reported for $\beta$-Ga$_2$O$_3$ \cite{Guo2015,Fu2018,Kumar2020}. 
On the other hand, the electron effective mass in $\beta$-Ga$_2$O$_3$ was found nearly isotropic from optical Hall effect measurements, in agreement with first principles calculations \cite{Mock2017,Furtmuller2016,Yamaguchi2004,Kang2017,Knight2018}. 

Of particular interest is the ability to characterize epitaxial materials since such are required for advanced electronic device structures. Depending on the substrate, epitaxial films may often be strained, which affects the material properties. 
Linear deformation-potential parameters have been derived from density-functional theory that can describe strain-related shift of phonon frequencies and band transitions for $\beta$-Ga$_2$O$_3$ \cite{Korlacki2020,Korlacki2022,KorlackiPRAppl2022}. 

Many epitaxial methods are currently employed for preparation of $\beta$-Ga$_2$O$_3$ and related materials. Homoepitaxial growth has been reported with great success for high structural quality of films. However, $\beta$-Ga$_2$O$_3$ single-crystal substrates are still cost intensive compared to, for example, sapphire. Homoepitaxial growth so far is necessary to meet the high requirements for lateral and vertical device architectures, but alternate substrates such as sapphire are being explored as well. Sapphire has been widely used in group-III nitride heteroepitaxy and is suitable for growth of $\beta$-Ga$_2$O$_3$ due to availability and similar template properties. 
Typically, $\beta$-Ga$_2$O$_3$ epitaxial layers with ($\bar{2}$01) orientation grow on \textit{c}-plane (0001) sapphire \cite{Huang2007,Seiler2015,Nakagomi2012,Lv2012,Gottschalch2009,Gogova2022}. 
Film textures and epitaxial relationships with the (0001) sapphire were studied for films grown by various methods such as pulsed laser deposition \cite{Seiler2015}, sputtering \cite{Lee2020}, gallium evaporation in oxygen plasma \cite{Nakagomi2012,Mochalov2020}, metal-organic chemical vapor deposition (MOCVD) 
 \cite{Lv2012,Gottschalch2009,Gogova2022,Gogova2015,Tadjer2016,Zhuo2017,Gottschalch2019}, halide vapor phase epitaxy \cite{Yao2018,Pozina2020,Mochalov2020a}, or molecular beam epitaxy \cite{Villora2006,Wei2019,Feng2021}. 

When $\beta$-Ga$_2$O$_3$ is heteroepitaxially grown on $c$-plane sapphire, six rotational domains with crystallographic ($\bar{2}$01) orientation are often revealed in X-ray diffraction (XRD) \cite{Huang2007,Seiler2015,Nakagomi2012,Lv2012,Gottschalch2009,Gogova2022}. 
They are rotated azimuthally 60$^\circ$ about the surface normal to the ($\bar{2}$01)-plane which is parallel to the $c$-axis of the (0001) sapphire substrate. 
$\beta$-Ga$_2$O$_3$ growth on mis-cut $c$-plane sapphire facilitates step flow growth mode and results in a reduced number of rotational domains \cite{Ma2022}. 
In particular, sapphire substrates with small mis-cut towards the $a$-face reduce the number of predominant rotation domains to three \cite{Gogova2022,Oshima2015}.  
For large mis-cut angles, reduction to mostly one domain can be achieved \cite{Oshima2015,Rafique2018}. 
If the substrate mis-cut is directed towards the $m$-face, all six rotation domains remain present. 

Developing methodologies for analysis of physical properties such as strain, bandgap energy, free charge carrier density, etc., are highly desirable for multiple domain ($\bar{2}$01) $\beta$-Ga$_2$O$_3$ films on $c$-plane sapphire. 
In this paper, we study the phonon mode properties of such films grown by MOCVD. 
Our specific aim is to provide an approach to correctly model the complex optical anisotropy in the infrared and far-infrared spectral regions taking into account the monoclinic structure of the individual crystal grains in the multi-domain films. 
The optical response of the films investigated here is best described by optically uniaxial film properties caused by  the highly symmetric arrangement of the rotation domains relative to the $c$-axis of sapphire. Upon comparison of our effective uniaxial model with experimental data during best-match model calculations of the measured infrared and far-infrared spectroscopic ellipsometry data, we obtain all TO and LO phonon mode parameters which are anticipated from bulk $\beta$-Ga$_2$O$_3$ investigations.\cite{Schubert2016}. 
We discuss the findings in view of the previously derived phonon mode model for single crystal monoclinic $\beta$-Ga$_2$O$_3$. 
We find deviations which we suggest here could be due to very small grain sizes of the multiple ($\bar2$01) oriented domains within the $\beta$-Ga$_2$O$_3$ films, and effects of regular arrangements of domains over length scales larger than the corresponding mode wavelengths (long range domain ordering). 
We believe that our model for the effective dielectric function provided in this work for the multiple domain $\beta$-Ga$_2$O$_3$ films can inspire development of future infrared and far-infrared analyses of films with multiple domains of low-symmetry materials using ellipsometry techniques.

\section{Theory}

\begin{figure}[!tbp]
  \begin{center}
 \includegraphics[width=1\linewidth]{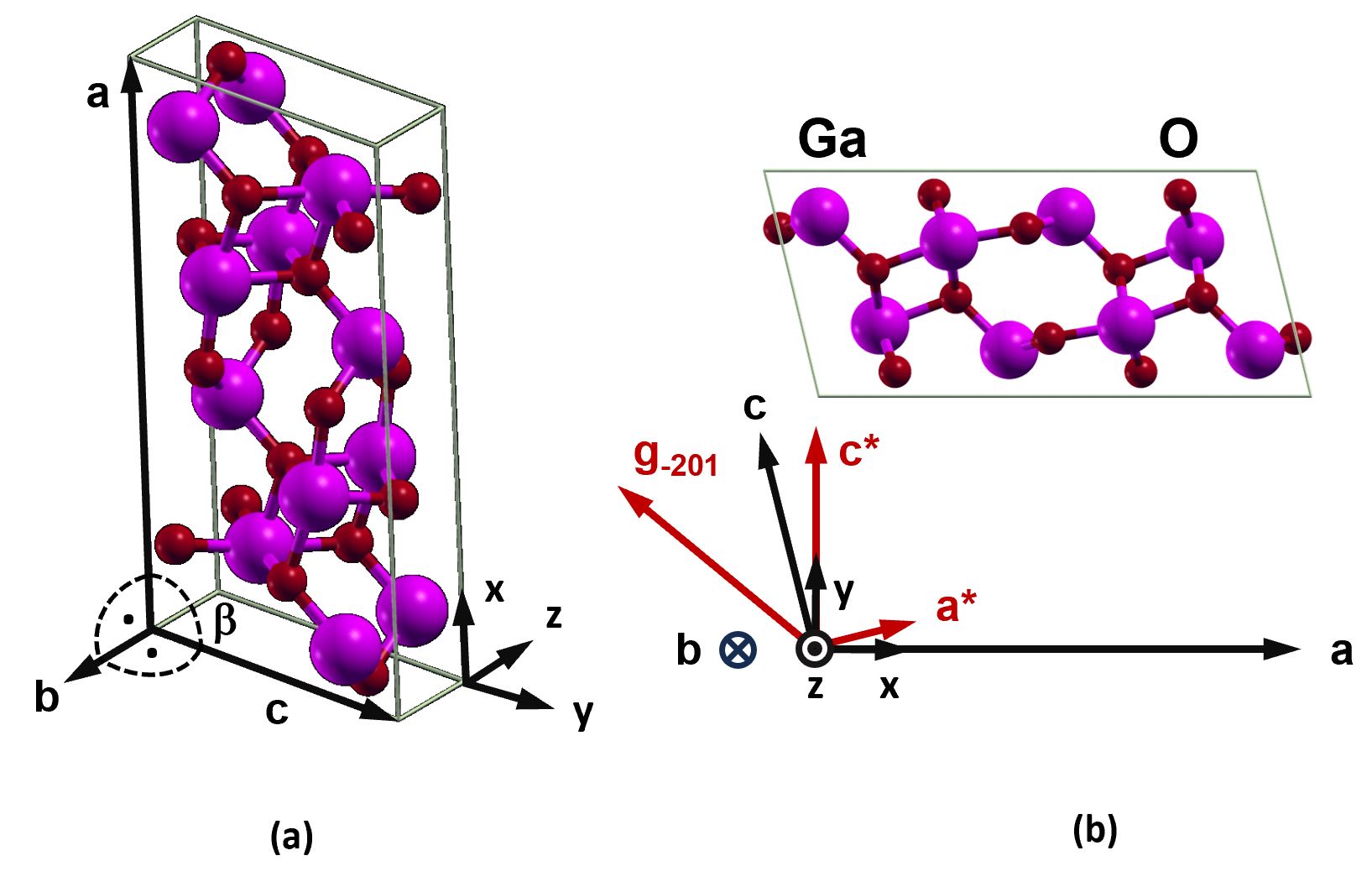}
    \caption{(a) Definition of the Cartesian laboratory coordinate system ($x$, $y$, $z$), and the unit cell of $\beta$-Ga$_2$O$_3$ with monoclinic angle $\beta$, and crystal unit vectors $\mathbf{a}$, $\mathbf{b}$, $\mathbf{c}$. (b) Monoclinic plane $\mathbf{a}$ - $\mathbf{c}$ viewed along vector $\mathbf{b}$. ($\mathbf{b}$ points into the plane.) Reciprocal vectors $\mathbf{a^{\star}}$, $\mathbf{c^{\star}}$ and $\mathbf{g_{\bar{2}01}}$ are shown for convenience.}
    \label{fig:Ga2O3unitcell}
  \end{center}
\end{figure}

\subsection{Infrared dielectric function model for monoclinic $\beta$-Ga$_2$O$_3$}

The unit cell of $\beta$-Ga$_2$O$_3$ is depicted in Fig.~\ref{fig:Ga2O3unitcell} with unit cell vectors $\mathbf{a}$, $\mathbf{b}$, and $\mathbf{c}$, and Cartesian coordinates $(x,y,z)$. A complete rendering for the dielectric tensor of $\beta$-Ga$_2$O$_3$ is needed, and was described in detail in Ref.~\cite{Schubert2016}. Three coordinate systems are required. Briefly, the first, $(a, b, c)$, addresses the crystal coordinates of the monoclinic structure, the second, $(\hat{x}, \hat{y}, \hat{z})$, implements the laboratory coordinate system tied to the ellipsometer instrument, the third, $(x, y, z)$, expresses the crystal coordinates in Cartesian coordinates, tied to a given choice relative to $(a, b, c)$, specifically in Fig.~\ref{fig:Ga2O3unitcell}, $(a,c^{\star},-b)$. A sample's normal direction is parallel to $\hat{z}$, which points into the surface of the sample.\cite{PhysRevB.53.4265} Plane $(\hat{x}-\hat{y})$ intersects $\hat{z}$ at its origin and is oriented parallel to the sample surface. Plane $(\hat{x}-\hat{z})$ is the plane of incidence for our ellipsometric measurements.

The dielectric tensor ($\varepsilon$) was described previously\cite{Schubert2016} using an eigendielectric displacement vector summation approach. Within $(x,y,z)$ coordinates in Fig.~\ref{fig:Ga2O3unitcell}

\begin{equation}
\varepsilon=\varepsilon_\infty+\sum_{l=1}^{12} \frac{A_{\mathrm{TO},l}^2}{\omega_{\mathrm{TO},l}^2-\omega^2-i\gamma_{\mathrm{TO},l}\omega}\,\mathbf{\hat{e}}_{\mathrm{TO},l}\otimes\mathbf{\hat{e}}_{\mathrm{TO},l}^\dagger.
\label{eq:Eigenpolarizationsummation}
\end{equation}

Symbol $\otimes$ indicates the Kronecker product operation which results in a dyadic that represents the contribution of mode $l$ in tensor form. Symbol $\dagger$ indicates the transpose and complex-conjugate operation such that the resulting dyadic is Hermitian or self-adjoint, and $i^2=-1$. The sum is taken over $l$ contributions from 12 infrared active optical phonon mode branches at the center of the Brillouin zone. Of these, 8 have $B_{\mathrm{u}}$ character and are polarized within the $(\mathbf{a}-\mathbf{c})$ plane. Four mode pairs with $A_{\mathrm{u}}$ character are polarized parallel to $\mathbf{b}$. All branches are associated with a transverse-optical (TO) and a longitudinal-optical (LO) mode. In the above equation, harmonically-broadened Lorentzian oscillators are used to model the dielectric response, and $A_{\mathrm{TO},l}$, $\omega_{\mathrm{TO},l}$, and $\gamma_{\mathrm{TO},l}$ then correspond to the amplitude, frequency, and broadening parameters in the so-called TO-summation form, respectively.\footnote{Note that an equivalent summation form can be established for the inverse dielectric function tensor thereby defining equivalent parameters for the LO-summation form. See, for example, the infrared phonon mode analysis reported for monoclinic yttrium orthosilicate by Mock~\textit{et al.}\cite{MockPRB2018YSO}. However, the considerations made in this present work for building the effective dielectric function tensor by averaging over contributions from multiple domains is not valid for the inverse dielectric function tensor summation forms. This is because the inverse of the sum of individual tensors is not equal to the sum of the inverses of individual tensors.} Unit vector $\mathbf{\hat{e}}_{\mathrm{TO},l}$ represents the dielectric displacement direction of TO mode $l$. $\varepsilon_\infty$ is the high-frequency dyadic contribution (see also Ref.\,\cite{Schubert2016,Schubert2020}). The resulting tensor has the following form within the coordinate system defined in Fig.~\ref{fig:Ga2O3unitcell}

\begin{equation}
\varepsilon = \begin{pmatrix} \varepsilon_{xx} & \varepsilon_{xy} & 0 \\ \varepsilon_{xy} & \varepsilon_{yy} & 0 \\ 0 & 0 & \varepsilon_{zz} \end{pmatrix}.
	\label{eq:epsilon010}
\end{equation}

\subsection{Euler rotations}

\begin{figure}[!tbp]
  \begin{center}
 \includegraphics[width=0.5\linewidth]{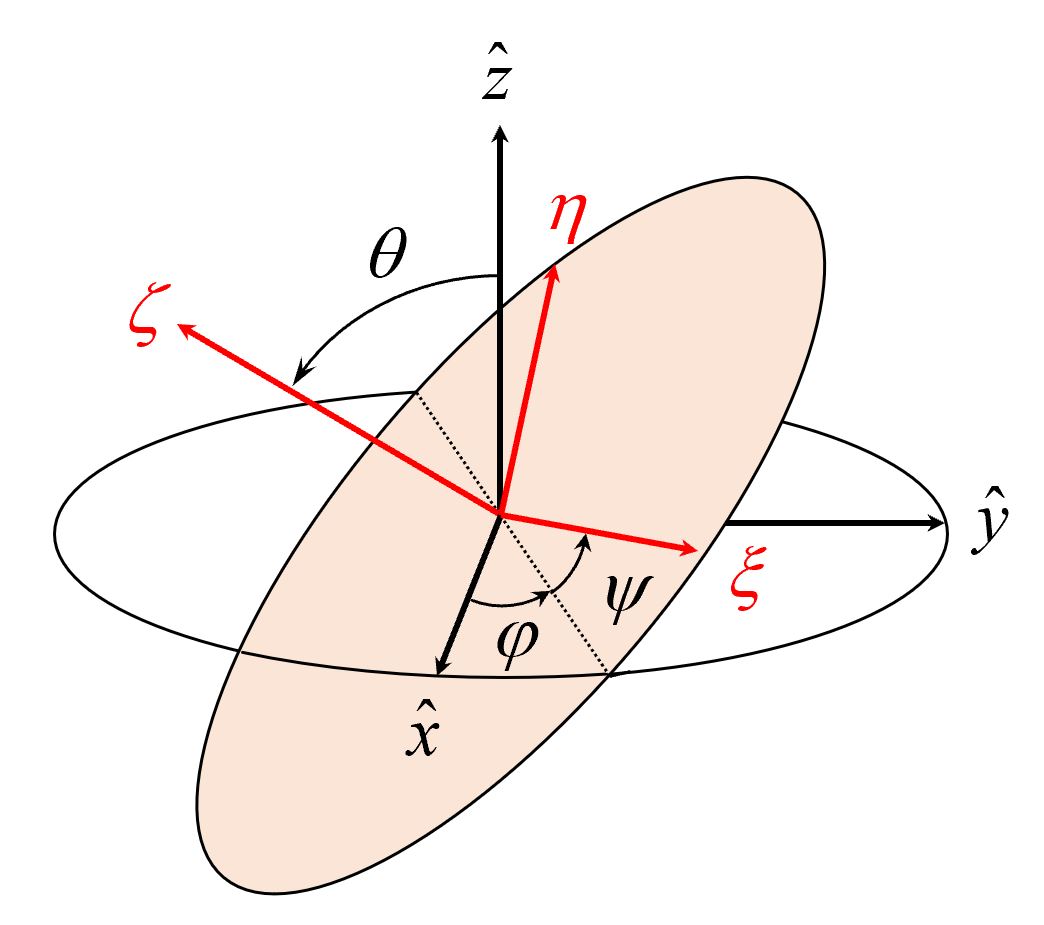}
    \caption{Definition of the Euler angles $\varphi$, $\theta$, and $\psi$ and the orthogonal rotations as provided by $\mathbf{A}$. ($\xi, \eta, \zeta $), and ($\hat{x}, \hat{y}, \hat{z}$) refer to the Cartesian auxiliary and laboratory coordinate systems, respectively. The auxiliary system here is (x,y,z). Reprinted from Ref.~\cite{Schubert2016} with copyright permission by American Physical Society.}
    \label{fig:Euler}
  \end{center}
\end{figure}

For a given sample, Euler angle rotations are required to bring $\varepsilon$ into the correct appearance within the ellipsometer system $(\hat{x}, \hat{y}, \hat{z})$. Figure~\ref{fig:Euler} depicts the Euler rotation operations used in this work. Two matrices, $R_1(v)$ and $R_2(v)$, are needed to rotate around $\hat{z}$ and $\hat{x}$ axes, respectively, with mathematically positive (negative) directions for positive (negative) arguments

\begin{equation}
R_1(v)=
\left(\begin{array}{ccc}
\cos v & -\sin v & 0 \\
\sin v &  \cos v & 0 \\
0 & 0 & 1
\end{array}\right),
\end{equation}
\begin{equation}
R_2(v)=
\left(\begin{array}{ccc}
1 & 0 & 0 \\
0 & \cos v & -\sin v \\
0 & \sin v &  \cos v
\end{array}\right).
\end{equation}

The full set of rotations, $\varphi, \theta, \psi$ indicated in Fig.~\ref{fig:Euler}, is then described by matrix $A$

\begin{equation}
A=R_1(\varphi)R_2(\theta)R_1(\psi),
\end{equation}

\noindent where $\hat{\varepsilon}$ indicates the tensor appearance of $\varepsilon$ in a new auxiliary system

\begin{equation}
\hat{\varepsilon}=A\varepsilon A^{-1}.
\end{equation}

\subsection{The six-domain $(\bar{2}01)$ model}\label{subsec:6domainmodel}

For a sample of $\beta$-Ga$_2$O$_3$ with the $(\bar{2}01)$ plane as its surface, the surface normal vector is the reciprocal vector $\mathbf{g_{\bar{2}01}}=-2\mathbf{a}^{\star}+\mathbf{c}^{\star}$, where $\mathbf{a}^{\star}$ and $\mathbf{c}^{\star}$ are reciprocal lattice vectors to $\mathbf{a}$ and $\mathbf{c}$, respectively (Fig.~\ref{fig:Ga2O3unitcell}b). $\mathbf{b}$ is parallel to the surface. The angle between $\mathbf{g_{\bar{2}01}}$ and vector $\mathbf{a}$ is $140^{\circ}$. Hence, Euler rotations $\theta=90^{\circ}$ and $\psi=50^{\circ}+90^{\circ}=130^{\circ}$ create the appearance of $\hat{\varepsilon}_{\bar{2}01}$ when such sample is oriented with its monoclinic plane lined up with the plane of incidence ($\varphi=0^{\circ}$).

\begin{equation}
\hat{\varepsilon}_{\bar{2}01}(\varphi=0^{\circ})=
\left(\begin{array}{ccc}
\hat{\varepsilon}_{xx} & 0 & \hat{\varepsilon}_{xy} \\
0 & \varepsilon_{zz} & 0 \\
\hat{\varepsilon}_{xy} & 0 & \hat{\varepsilon}_{yy}
\end{array}\right),\label{eq:domain0}
\end{equation}

\noindent where $\hat{\varepsilon}_{xx}$, $\hat{\varepsilon}_{xy}$, $\hat{\varepsilon}_{yy}$ are obtained from $\varepsilon_{xx}$, $\varepsilon_{xy}$, $\varepsilon_{yy}$ in  Eq.~\ref{eq:epsilon010} by rotation within the monoclinic plane using $\psi=130^{\circ}$, and $\varepsilon_{zz}$ is the same as in  Eq.~\ref{eq:epsilon010}. For the specific sample configuration discussed in this work, six equally weighted-by-volume domains exist within an epitaxial layer. Each domain is rotated around its surface normal, $\mathbf{g_{\bar{2}01}}$, by $\varphi=0^{\circ}$, $60^{\circ}$, $120^{\circ}$, $180^{\circ}$, $240^{\circ}$, or $300^{\circ}$, respectively. In a linear summation approximation, the effective dielectric function tensor of such epitaxial layer can be expressed as a sum over the dielectric function tensor of each domain weighed in equal parts

\begin{equation}
\hat{\varepsilon}_\text{eff}
=\frac{1}{6}\sum_{\substack{\varphi=0,\,60,\,120,\\180,\,240,\,300^\circ}}R_1(\varphi)\hat{\varepsilon}_{(\bar{2}01)}R^{-1}_1(\varphi),
\label{eq:6domainsum}
\end{equation}

\noindent where the terms appear as follows

\begin{equation}
\hat{\varepsilon}_{\bar{2}01}(60^{\circ})=
\left(\begin{array}{ccc}
\frac{(\hat\varepsilon_{xx}+3\varepsilon_{zz})}{4} & \frac{\sqrt{3}(\hat\varepsilon_{xx}-\varepsilon_{zz})}{4} & \frac{\hat\varepsilon_{xy}}{2} \\
\frac{\sqrt3(\hat\varepsilon_{xx}-\varepsilon_{zz})}{4} & 
\frac{(3\hat\varepsilon_{xx}+\varepsilon_{zz})}{4} & \frac{\sqrt{3}\hat\varepsilon_{xy}}{2} \\
\frac{\hat\varepsilon_{xy} }{2}& \frac{\sqrt{3}\hat\varepsilon_{xy}}{2} & \hat\varepsilon_{yy}
\end{array}\right),
\end{equation}

\begin{equation}
\hat{\varepsilon}_{\bar{2}01}(120^{\circ})=
\left(\begin{array}{ccc}
\frac{(\hat\varepsilon_{xx}+3\varepsilon_{zz})}{4} & \frac{\sqrt{3}(-\hat\varepsilon_{xx}+\varepsilon_{zz})}{4} & -\frac{\hat\varepsilon_{xy}}{2} \\
\frac{\sqrt3(-\hat\varepsilon_{xx}+\varepsilon_{zz})}{4} & 
\frac{(3\hat\varepsilon_{xx}+\varepsilon_{zz})}{4} & \frac{\sqrt{3}\hat\varepsilon_{xy}}{2} \\
-\frac{\hat\varepsilon_{xy} }{2}& \frac{\sqrt{3}\hat\varepsilon_{xy}}{2} & \hat\varepsilon_{yy}
\end{array}\right),
\end{equation}

\begin{equation}
\hat{\varepsilon}_{\bar{2}01}(180^{\circ})=
\left(\begin{array}{ccc}
\hat\varepsilon_{xx} & 0 & -\hat\varepsilon_{xy} \\
0 & \varepsilon_{zz} & 0 \\
-\hat\varepsilon_{xy}& 0 & \hat\varepsilon_{yy}
\end{array}\right),\label{eq:domain180}
\end{equation}

\begin{equation}
\hat{\varepsilon}_{\bar{2}01}(240^{\circ})=
\left(\begin{array}{ccc}
\frac{(\hat\varepsilon_{xx}+3\varepsilon_{zz})}{4} & \frac{\sqrt{3}(\hat\varepsilon_{xx}-\varepsilon_{zz})}{4} & -\frac{\hat\varepsilon_{xy}}{2} \\
\frac{\sqrt3(\hat\varepsilon_{xx}-\varepsilon_{zz})}{4} & 
\frac{(3\hat\varepsilon_{xx}+\varepsilon_{zz})}{4} & -\frac{\sqrt{3}\hat\varepsilon_{xy}}{2} \\
-\frac{\hat\varepsilon_{xy} }{2}&- \frac{\sqrt{3}\hat\varepsilon_{xy}}{2} & \hat\varepsilon_{yy}
\end{array}\right),
\end{equation}

\begin{equation}
\hat{\varepsilon}_{\bar{2}01}(300^{\circ})=
\left(\begin{array}{ccc}
\frac{(\hat\varepsilon_{xx}+3\varepsilon_{zz})}{4} & \frac{\sqrt{3}(-\hat\varepsilon_{xx}+\varepsilon_{zz})}{4} & \frac{\hat\varepsilon_{xy}}{2} \\
\frac{\sqrt3(-\hat\varepsilon_{xx}+\varepsilon_{zz})}{4} & 
\frac{(3\hat\varepsilon_{xx}+\varepsilon_{zz})}{4} &- \frac{\sqrt{3}\hat\varepsilon_{xy}}{2} \\
\frac{\hat\varepsilon_{xy} }{2}& -\frac{\sqrt{3}\hat\varepsilon_{xy}}{2} & \hat\varepsilon_{yy}
\end{array}\right).
\end{equation}

Hence, after summation

\begin{equation}
\hat{\varepsilon}_{\mathrm{eff}}=
\left(\begin{array}{ccc}
\varepsilon_{\perp} & 0 & 0 \\
0 & \varepsilon_{\perp} & 0 \\
0 & 0 & \varepsilon_{||}
\end{array}\right),
\end{equation}

\noindent where we defined effective in-plane and out-of-plane dielectric functions, $\varepsilon_{\perp}$ and $\varepsilon_{||}$, respectively

\begin{equation}
\varepsilon_{\perp}=\frac{1}{2}(\hat\varepsilon_{xx}+\varepsilon_{zz}),
\end{equation}
\begin{equation}
\varepsilon_{||}=\hat\varepsilon_{yy}.
\end{equation}

We obtain thereby an effective dielectric tensor which renders the optical response of an effectively optically uniaxial material, with ordinary and extraordinary optical constants, $n_{\perp,||}+ik_{\perp,||}=\sqrt{\varepsilon_{\perp,||}}$. The optic axis is oriented along surface normal $\mathbf{g}_{\bar{2}01}$. No coupling of $p$ into $s$ polarized wave components and vice versa upon reflection or transmission is anticipated or observed. Therefore, standard ellipsometry measurements are sufficient and measurements of multiple sample orientations is unnecessary. Note that $\hat\varepsilon_{xx}$ is obtained when all unit vectors $\hat{\mathbf{e}}_{\mathrm{TO},l}$ in Eq.~\ref{eq:Eigenpolarizationsummation} are shifted by $130^{\circ}$ according to the rotation within the monoclinic plane by $130^{\circ}$. Then, all TO modes with $B_{\mathrm{u}}$ character cause signatures in $\hat\varepsilon_{xx}$. All modes with $A_{\mathrm{u}}$ character cause signatures in $\varepsilon_{zz}$.  Hence, all TO modes are anticipated to cause signatures in $\varepsilon_{\perp}$. However, $\varepsilon_{||}$ only consists of $\hat\varepsilon_{yy}$, and which only contains contributions from TO modes with $B_{\mathrm{u}}$ character. Hence, for $\varepsilon_{||}$ only signatures due to modes with $B_{\mathrm{u}}$ character are anticipated. Further, the amplitudes of TO modes with $B_{\mathrm{u}}$ character in $\varepsilon_{\perp}$ are expected according to their projection onto the sample surface $(\bar{2}01)$. Due to this, modes that have a small projection may cause limited or no signatures. Likewise, modes which have large projection onto the sample normal $\mathbf{g}_{\bar{2}01}$ will have strong contributions to $\varepsilon_{||}$.

\subsection{The effective tensor model dielectric function}\label{sec:effectiveuniaxialtensormodel}
\subsubsection{Effective TO mode summation form}
Instead of using the eigendielectric displacement vector summation approach, we suggest to introduce another TO form summation approach for $\hat{\varepsilon}_\text{eff}$. Thereby, we introduce sums of harmonically broadened Lorentizan oscillators which permit us to quantitatively determine TO phonon mode parameters

\begin{align}
	\varepsilon_\perp = \varepsilon_{\infty,\perp}+\sum_{l=1}^{8+4} \frac{A_{\text{TO},\perp,l}^2}{\omega_{\mathrm{TO},l}^2-\omega^2-i\gamma_{\mathrm{TO},l}\omega} \,, \label{eq:uniaxialDFcomponents1}\\
	\varepsilon_{||} = \varepsilon_{\infty,||}+\sum_{l=1}^{8} \frac{A_{\mathrm{TO},||,l}^2}{\omega_{\mathrm{TO},l}^2-\omega^2-i\gamma_{\mathrm{TO},l}\omega}\,,
	\label{eq:uniaxialDFcomponents2}
\end{align}

\noindent with amplitude, TO frequency, and TO broadening parameters $A_{\text{TO},\perp;||,l}$, $\omega_{\text{TO},\perp;||,l}$, and $\gamma_{\text{TO},\perp;||,l}$, for ordinary ($\perp$) and extraordinary $(||)$ components. Note that this effective uniaxial dielectric function model can also be applied for band gap modeling of such films \cite{Gogova2022}. 

\subsubsection{Effective LO mode summation form}

The same summation approach can be applied to the inverse effective dielectric functions~\cite{Mock2019NGO}

\begin{align}
	\varepsilon^{-1}_\perp = \varepsilon^{-1}_{\infty,\perp}-\sum_{l=1}^{8+4} \frac{A_{\text{LO},\perp,l}^2}{\omega_{\mathrm{LO},l}^2-\omega^2-i\gamma_{\mathrm{LO},l}\omega} \,, \label{eq:uniaxialinvDFcomponents1}\\
	\varepsilon_{||} = \varepsilon^{-1}_{\infty,||}-\sum_{l=1}^{8} \frac{A_{\mathrm{LO},||,l}^2}{\omega_{\mathrm{LO},l}^2-\omega^2-i\gamma_{\mathrm{LO},l}\omega}\,,
	\label{eq:uniaxialinvDFcomponents2}
\end{align}

\noindent with LO amplitude, LO frequency, and LO broadening parameters $A_{\text{LO},\perp;||,l}$, $\omega_{\text{LO},\perp;||,l}$, and $\gamma_{\text{LO},\perp;||,l}$, for ordinary ($\perp$) and extraordinary $(||)$ inverse dielectric function tensor elements. Note the minus sign in front of the summation, which ensures real-valued parameters for amplitudes\cite{Mock2019NGO}. Note that~Eqs.~\ref{eq:uniaxialDFcomponents1},~\ref{eq:uniaxialDFcomponents2} and ~Eqs.~\ref{eq:uniaxialinvDFcomponents1},~\ref{eq:uniaxialinvDFcomponents2} establish two alternative dispersion models. Having a full set of parameters to Eqs.~\ref{eq:uniaxialDFcomponents1},\ref{eq:uniaxialDFcomponents2}, in principle, determines all parameters in Eqs.~\ref{eq:uniaxialinvDFcomponents1},\ref{eq:uniaxialinvDFcomponents2}, and vice versa. Using the different dispersion model approaches during the analysis of experimental ellipsometry data provides added sensitivity to the two parameter sets.

\section{Experimental details and methods}

We have studied $\beta$-Ga$_2$O$_3$ films with typical six-fold rotation domains grown by MOCVD on $c$-plane sapphire. No miscut was utilized in the substrates investigated here. Details on $\beta$-Ga$_2$O$_3$ growth and properties can be found elsewhere \cite{Lv2012,Gogova2014,Gogova2015,BinAnooz2021}. 

The samples were investigated by XRD (Philips X'Pert MRD and Panalytical Empyrean) to confirm the $(\bar{2}01)$-plane orientation by measurements in Bragg-Brentano geometry. Pole figures of various crystal planes ($(\bar{2}01)$, (001)/(002), (111), (400)) were obtained to confirm the equal distribution of rotation domains. The results for all films are very similar independent of the film thickness and specific growth conditions. A representative pole figure is presented in Fig.~\ref{fig:PF} for the (001) Bragg reflection. The lattice parameters ($a,b,c$) and the angle $\beta$  were determined from the 2$\theta - \omega$ positions of the $\beta$-Ga$_2$O$_3$ $\bar{2}01$, 002, 400 and 111 XRD peaks for all films. The  strain values along the main crystallographic directions were estimated from the measured lattice parameters and the respective lattice parameters and angle $\beta$ of an undoped $\beta$-Ga$_2$O$_3$ $(\bar{2}01)$ oriented single crystal.  All $\beta$-Ga$_2$O$_3$ $(\bar{2}01)$ films were found to have similar levels of strain.



Mid- and far-infrared ellipsometry using a commercial (J.~A.~Woollam~Co.,~Inc., IR-VASE) and a home-built ellipsometer~\cite{Kuehne2014,Kuhne2018}, respectively, were performed. Both instruments are based on Fourier-transform infrared spectroscopy. While the first one employs a rotatable compensator for the measurements, the latter is a rotating-polarizer rotating-analyzer configuration ellipsometer. Hence, the first three columns of the 4$\times$4 Mueller matrix are accessible in the mid-infrared range while only the upper left $3\times3$ sub-matrix is available in the far-infrared (see Fig.\,\ref{fig:SE} below). Hence, the Mueller matrix elements in the fourth row are only presented in the spectral range $\approx$250-900~cm$^{-1}$. The combined spectral ranges of both ellipsometers cover all optical phonon mode frequencies of $\beta$-Ga$_2$O$_3$. 

The optical properties of the sapphire substrate were calculated according to Schubert, Herzinger and Tiwald.~\cite{PhysRevB.61.8187}. The Berreman formalism in the extension described by Schubert was used for calculation of the ellipsometric parameters for the best-match model analysis.\cite{Berreman:72,PhysRevB.53.4265} The effective dielectric functions of the $\beta$-Ga$_2$O$_3$ films were modelled according to Eqs.\,\ref{eq:uniaxialDFcomponents1} and \ref{eq:uniaxialDFcomponents2}, and according to Eqs.\,\ref{eq:uniaxialinvDFcomponents1} and \ref{eq:uniaxialinvDFcomponents2} for the inverse effective dielectric functions. TO mode terms in $\varepsilon_{\perp}$ and $\varepsilon_{\parallel}$ with $B_{\mathrm{u}}$ character were assumed to share equal frequency and broadening parameters. Best-match calculations of infrared ellipsometry parameters matching the experimental data were obtained by first using the TO sum forms, and secondly by using the LO sum forms. The two resulting best-match calculated data are virtually indistinguishable. 

We also included an experimental-difference data best-match model approach. We calculated virtual, experimental-difference ellipsometry data, $\delta M_{ij}$, by subtracting from the measured data the model calculated ellipsometry data for the same wavelength and incidence conditions assuming a $c$-plane sapphire substrate only. We assumed the same uncertainty parameters determined from the measurement also for the $\delta M_{ij}$ data. These differences data were then included into the best-match model calculations. We thereby matched $M_{ij}$ and $\delta M_{ij}$ simultaneously. The model-calculated difference data were obtained by subtracting at every iteration step model calculated data for a bare substrate from the model calculated data which included the epitaxial layer into the optical model. The results, using the TO sum forms, are shown in Fig.~\ref{fig:SE} ($M_{ij}$) and Fig.~\ref{fig:SEdiff} ($\delta M_{ij}$). The rationale for including data $\delta M_{ij}$ into the model analysis is simply because these data instantly reveal the spectral regions and magnitudes of changes induced by the epitaxial layer, and thereby indicate how strong certain features are related to their associated phonon modes.

\section{Results and discussion}

\begin{figure}
	\centering
	\includegraphics{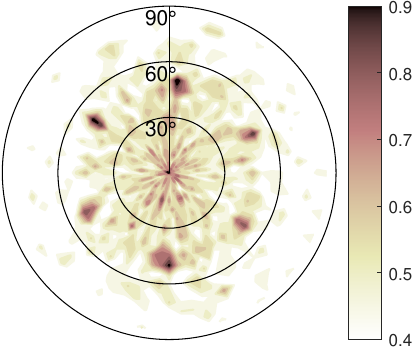}
	\caption{Pole figure of normalized XRD intensity for the (001) Bragg reflection of a 220\,nm thick $\beta$-Ga$_2$O$_3$ film on $c$-plane sapphire substrate. The six approximately equally distributed rotation domains are clearly visible. }
	\label{fig:PF}
\end{figure}

\begin{figure*}
	\centering
	\includegraphics[width=1\linewidth]{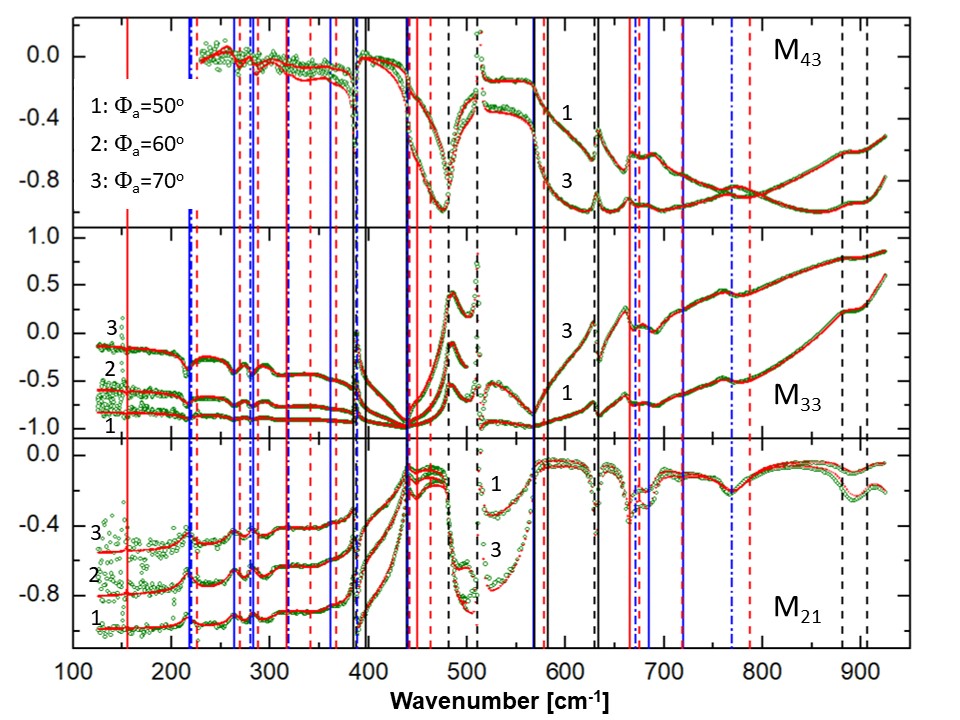}
	\caption{Experimental (green symbols) and best-matching-model calculated (red line using TO sum forms in Eqs.~\ref{eq:uniaxialDFcomponents1},\ref{eq:uniaxialDFcomponents2}) Mueller matrix element spectra of a 220\,nm thick $\beta$-Ga$_2$O$_3$ film on the $c$-plane sapphire substrate. Only block diagonal elements except for $M_{22}$ are shown. Block off-diagonal elements are zero. $M_{12}$ and $M_{21}$ are identical. Black vertical lines indicate sapphire TO (solid lines) and LO (dashed lines) modes with $A_{\mathrm{2u}}$ and $E_{\mathrm{u}}$ character, see also Ref.~\cite{PhysRevB.61.8187}. Solid vertical lines indicate film TO frequencies for electric field polarization parallel (blue) and perpendicular (red) to the sample surface. Dashed vertical lines indicate film LO frequencies for polarization parallel (blue) and perpendicular (red). Labels indicate the angle of incidence under which data were acquired.}
	\label{fig:SE}
\end{figure*}

\begin{figure*}
	\centering
	\includegraphics[width=1\linewidth]{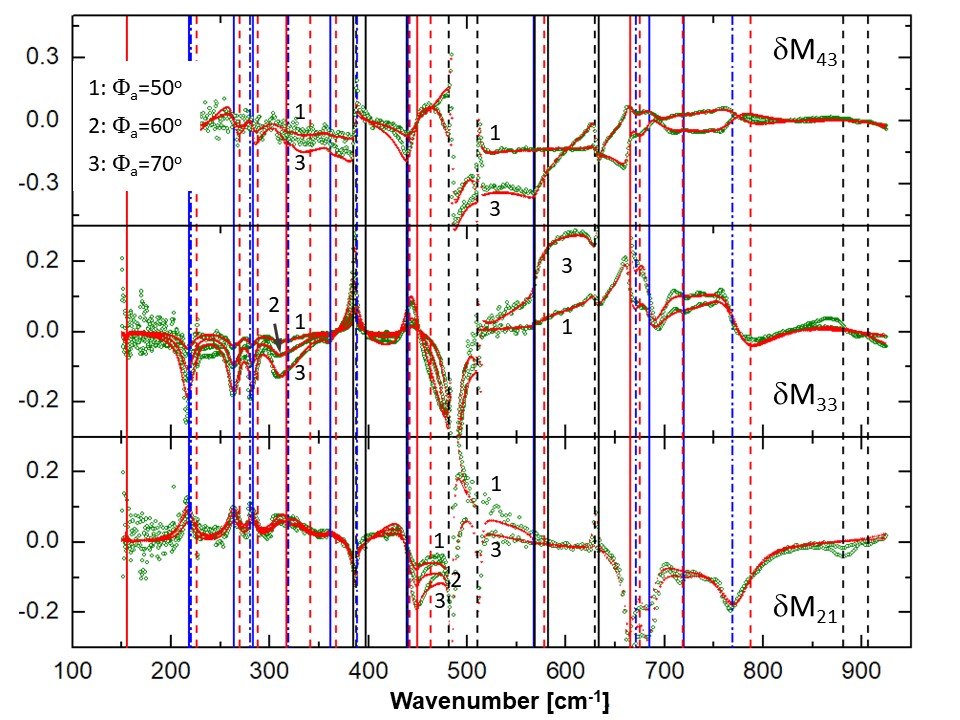}
	\caption{Same as Fig.~\ref{fig:SE} but for difference data $\delta M_{ij}$. The difference in Mueller matrix elements are obtained by subtracting experimental data (green symbols) and best-match model calculated data (red symbols) by model calculated Mueller matrix elements obtained for $c$-plane sapphire. The difference data are most sensitive to the epitaxial layer properties.}
	\label{fig:SEdiff}
\end{figure*}

The lattice mismatch between $\beta$-Ga$_2$O$_3$ and sapphire in case of pseudomorphic growth of $(\bar{2}01)$-oriented domains should result in compressive strain along $[102]$ and tensile strain along $[010]$. The proximity of differently rotated domains may result in additional biaxial strain within the $(\bar{2}01)$-plane.\cite{Chen2015} In all films the normal and shear strain levels were found to be similar and not exceeding 5x10$^-3$. We therefore consider our sample nearly relaxed. Relaxation was previously reported to initiate at the interface with the substrate by 3-4 atomic layers of $\upalpha$-Ga$_2$O$_3$ that grow pseudomorphic to sapphire.\cite{Gogova2022,Schewski2014} After this critical thickness, the structure transforms and continues to grow in a relaxed $\beta$-phase. Figure~\ref{fig:PF} shows a representative (001) XRD pole figure of a 220\,nm thick $\beta$-Ga$_2$O$_3$ film on $c$-plane oriented sapphire substrate. The pole figure reveals six rotational domains. Quantitative analysis of the six peak intensities was carried out and confirmed equal domain distribution with $16.7\pm1.5$  volume fraction percent XRD intensity contributed by each domain. 


Figure~\ref{fig:SE} shows measured and modeled ellipsometry spectra in terms of block on-diagonal Mueller matrix elements, $M_{ij}$. Figure~\ref{fig:SEdiff} shows the corresponding virtual experimental-difference ellipsometry data, $\delta M_{ij}$. The block off-diagonal elements are zero within the error bars, and are not shown. Also not shown are data measured at different in-plane sample azimuth orientations because these data are equal to the data shown within the error bar. Hence, no conversion from $p$ to $s$ polarization and vice versa upon reflection from the surface was observed regardless of the sample azimuth orientation. We conclude, in agreement with the results from XRD investigations, that the existence of equally distributed domains has averaged the monoclinic anisotropy of the $(\bar{2}01)$-$\beta$-Ga$_2$O$_3$ domains to an effective isotropic optical behavior within the film surface. We note that the angle of incidence dependence of the Mueller matrix data in Fig.~\ref{fig:SE} cannot be modelled with an optically isotropic model. Instead, the effective uniaxial model discussed in Sec.~\ref{sec:effectiveuniaxialtensormodel} must be used.  Good agreement between experimental data and the effective uniaxial model is found as shown in Fig.~\ref{fig:SE}. The resonance frequencies of the effective in-plane ($\perp$) and out-of-plane ($||$) dielectric function TO modes (Eqs.~\ref{eq:uniaxialDFcomponents1},~\ref{eq:uniaxialDFcomponents2}) and inverse dielectric function LO modes (Eqs.~\ref{eq:uniaxialinvDFcomponents1},~\ref{eq:uniaxialinvDFcomponents2}) are indicated by vertical lines in Fig.~\ref{fig:SE}. Additional lines indicate two pairs of TO and LO modes with $A_{2u}$ character (parallel sapphire substrate lattice direction $c$) and four pairs with $E_u$ character (perpendicular $c$) for sapphire where the latter appears here with infrared optical signatures due to its $c$-plane orientation.\cite{PhysRevB.61.8187} Note that Mueller matrix element $M_{43}$ is only shown within the range of our IR ellipsometer system, while our FIR ellipsometer system cannot measure this information.

\begin{figure}
	\centering
	\includegraphics[width=1\linewidth]{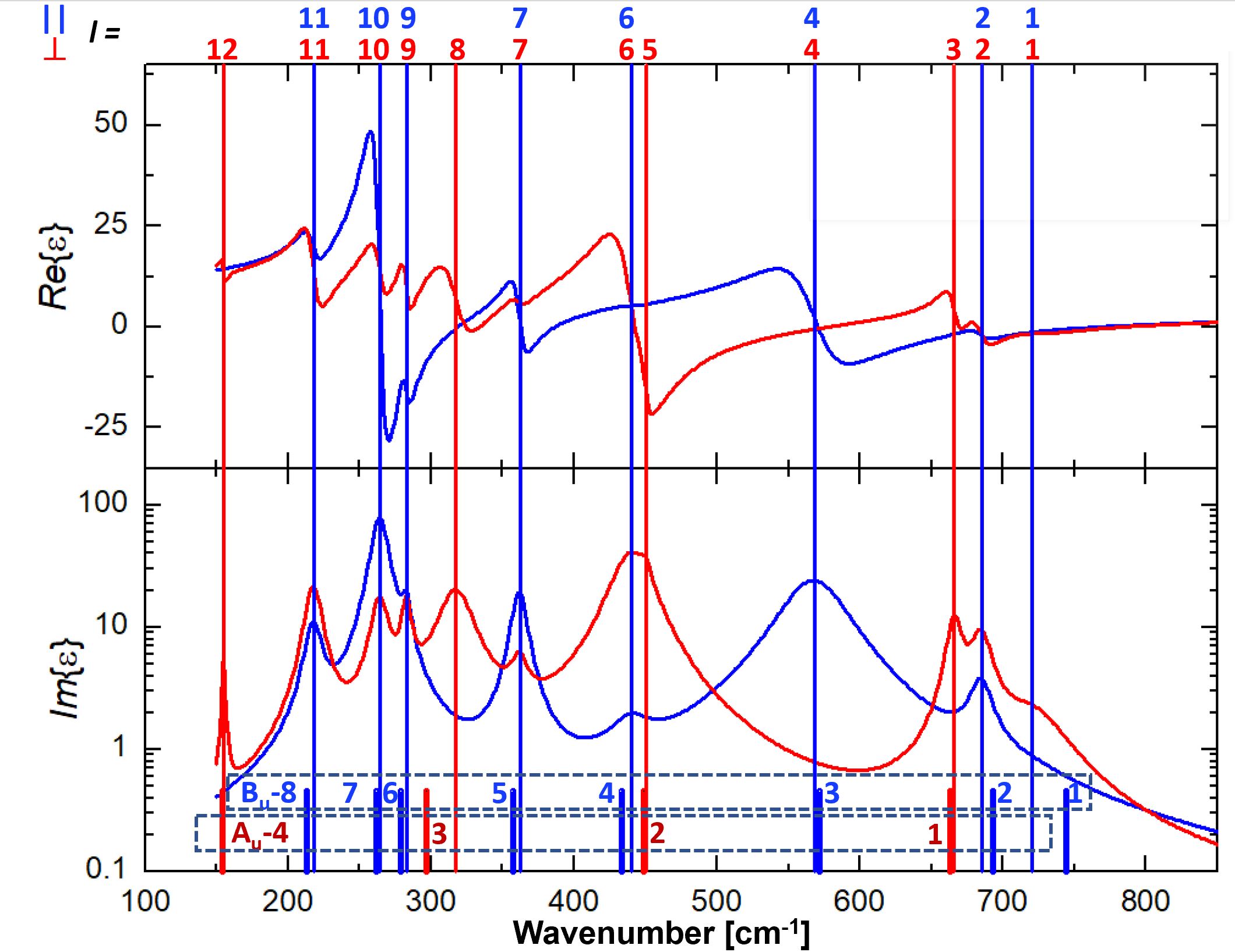}
	\caption{Real (top panel) and imaginary (bottom panel) effective uniaxial model dielectric function for ($\bar{2}01$)-plane oriented $\beta$-Ga$_2$O$_3$ films with equally distributed rotation domains for polarization parallel (blue) and perpendicular (red) to the surface normal (reciprocal vector $\mathbf{g}_{\bar{2}01}$). 12 TO mode frequencies are identified with index $l$, and indicated by vertical solid lines for polarization parallel (blue) and perpendicular (red). Also indicated are TO phonon mode frequencies with $A_{\mathrm{u}}$ (vertical short thick red solid lines) and $B_{\mathrm{u}}$ symmetry (vertical short thick blue solid lines) from previous investigations of single crystalline $\beta$-Ga$_2$O$_3$ (Ref.\,\cite{Schubert2019a}). Modes $l=1,2,4,6,7,9,10,11$ occur in both effective dielectric functions, and modes $l=3,5,8,12$ only occur for perpendicular polarization. The latter appear near frequencies of single crystal $A_{\mathrm{u}}$ modes, while the former appear near frequencies of $B_{\mathrm{u}}$ modes.}
	\label{fig:DF}
\end{figure}

\begin{figure}
	\centering
\includegraphics[width=1\linewidth]{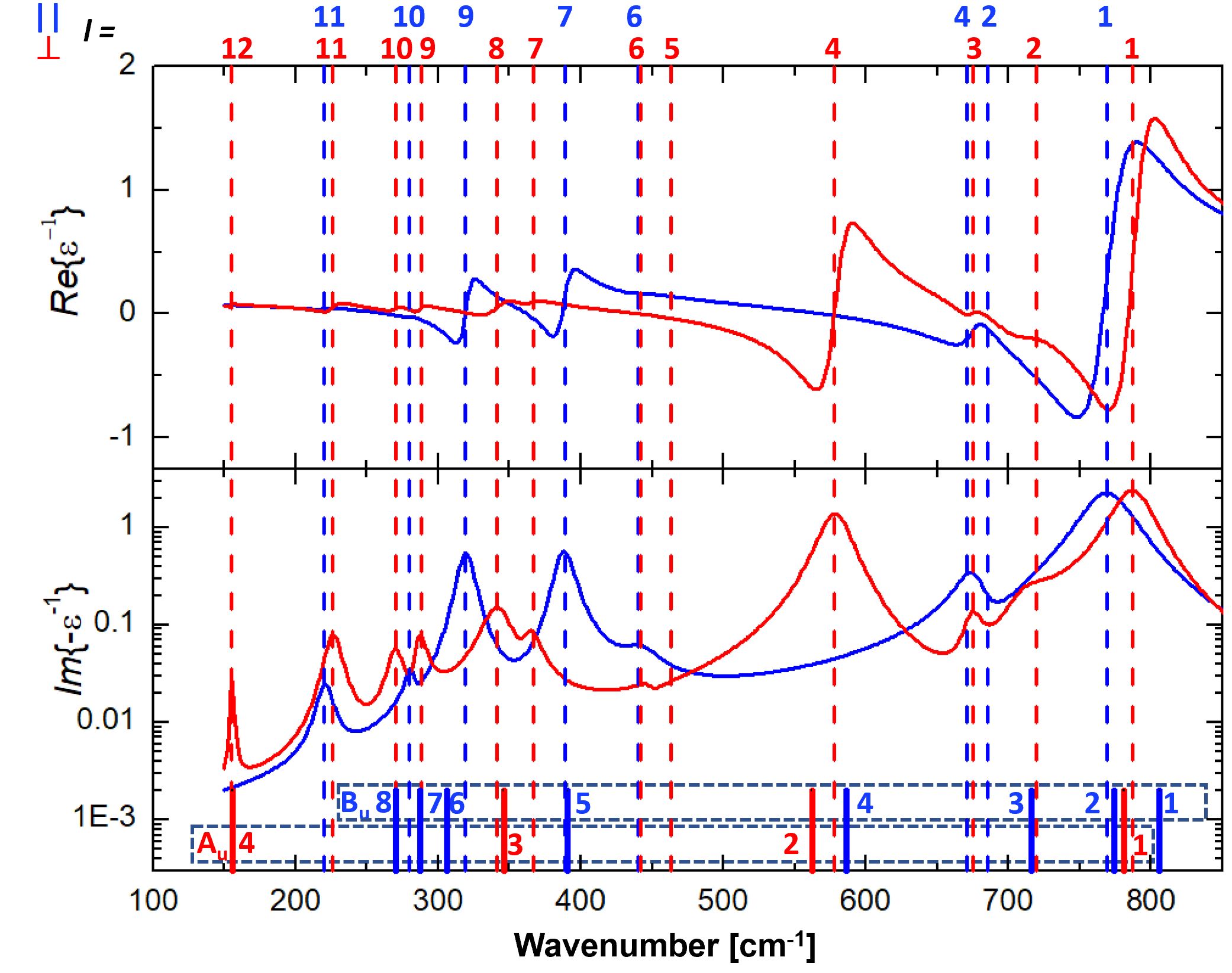}
	\caption{Same as Fig.~\ref{fig:DF} for inverse model dielectric functions depicting LO mode frequencies. LO mode frequencies for single crystalline $\beta$-Ga$_2$O$_3$ are included for comparison (Ref.\,\cite{Schubert2019a}). Modes observed in this work with index $l=1,2,4,6,7,9,10,11$  differ between parallel and perpendicular polarization (vertical dashed lines).}
	\label{fig:invDF}
\end{figure}

\begin{table*}
	\centering
	\caption{Effective in-plane ($\perp$) and out-of-plane ($||$) model dielectric function TO (Eqs.~\ref{eq:uniaxialDFcomponents1},~\ref{eq:uniaxialDFcomponents2}) and LO (Eqs.~\ref{eq:uniaxialinvDFcomponents1},~\ref{eq:uniaxialinvDFcomponents2}) sum form parameters from best-match model analysis of ellipsometry data shown in Fig.~\ref{fig:SE} taken from a six-fold $(\bar{2}01)$ domain thin film on $c$-plane sapphire. Parenthesis indicate the last significant digit within which the same parameters will be found in repeated iterations with 90\% probability. All parameters in units of cm$^{-1}$. Mode index $l$ corresponds to phonon modes in bulk $\beta$-Ga$_2$O$_3$. Modes of character $B_{\mathrm{u}}$ are projected both parallel and perpendicular to the film normal $\mathbf{g}_{\bar{2}01}$, modes of character $A_{\mathrm{u}}$ are only projected perpendicular to $\mathbf{g}_{\bar{2}01}$: $l=1: B_u-1$, $l=2: B_u-2$, $l=3: A_u-1$, $l=4: B_u-3$, $l=5: A_u-2$, $l=6: B_u-4$, $l=7: B_u-5$, $l=8: A_u-3$, $l=9: B_u-6$, $l=10: B_u-7$, $l=11: B_u-8$, $l=12: A_u-4$. The effective high-frequency dielectric constants are obtained from Eqs.~\ref{eq:uniaxialDFcomponents1},\ref{eq:uniaxialDFcomponents2} $\varepsilon_{\infty,\perp}=3.(6)$ and $\varepsilon_{\infty,||}=3.(7)$.}
\begin{tabular}{cccccccccccccc}
\hline \hline
 $l$&$M^{a}$&$A_{\mathrm{TO},\perp}$ & $\omega_{\mathrm{TO},\perp}$ & $\gamma_{\mathrm{TO},\perp}$ & $A_{\mathrm{LO},\perp}$ & $\omega_{\mathrm{LO},\perp}$ & $\gamma_{\mathrm{LO},\perp}$& $A_{\mathrm{TO},||}$ & $\omega_{\mathrm{TO},|| }$ & $\gamma_{\mathrm{TO},||}$& $A_{\mathrm{LO},||}$ & $\omega_{\mathrm{LO},||}$ & $\gamma_{\mathrm{LO},||}$ \\
		\hline \hline
1&$B_u$-1&25(6)&72(0)&5(5)&24(7)&78(7)&3(2)&(5)0&$^b$&$^b$&26(9)&76(9)&4(2)\\
2&$B_u$-2&30(0)&68(5)&1(7)&6(4)&71(9)&3(1)&17(5)&$^b$&$^b$&5(1)&68(5)&2(0)0\\
3&$A_u$-1&28(8)&66(6)&1(2)&1(7)&67(5)&(5)&$^c$&$^c$&$^c$&$^c$&$^c$&$^c$\\
4&$B_u$-3&(4)&56(8)&5(2)&14(1)&57(8)&2(5)&83(8)&$^b$&$^b$&5(7)&670.(7)&2(1)\\
5&$A_u$-2&18(8)&45(0)&(8)&(1)&46(3)&1(1)&$^c$&$^c$&$^c$&$^c$&$^c$&$^c$\\
6&$B_u$-4&74(1)&44(0)&3(2)&1(6)&44(2)&12(1)&10(6)&$^b$&$^b$&3(0)&440.(5)&4(2)\\
7&$B_u$-5&11(6)&36(2)&1(2)&1(6)&36(7)&1(5)&28(6)&$^b$&$^b$&57.(4)&388.(6)&15.(3)\\
8&$A_u$-3&40(2)&31(7)&2(6)&3(5)&34(1)&2(6)&$^c$&$^c$&$^c$&$^c$&$^c$&$^c$\\
9&$B_u$-6&16(4)&28(3)&(7)&1(4)&28(8)&1(1)&15(0)&$^b$&$^b$&46.(8)&31(9)&1(3)\\
10&$B_u$-7&23(1)&26(4)&1(3)&1(1)&27(0)&1(1)&52(1)&$^b$&$^b$&(4)&28(0)&(4)\\
11&$B_u$-8&24(1)&217.(8)&1(3)&1(5)&22(6)&1(5)&16(3)&$^b$&$^b$&(6)&22(0)&(9)\\
12&$A_u$-4&47$^d$&155$^d$&2.5$^d$&(4)&156$^d$&(3)&$^c$&$^c$&$^c$&$^c$&$^c$&$^c$\\
		\hline \hline
	\end{tabular}\\
 $^a$ Mode assignment according to irreducible presentation in single crystalline $\beta$-Ga$_2$O$_3$.\\
 $^b$ Same parameter as the perpendicular TO sum.\\
 $^c$ Transition dipole of phonon mode does not have component parallel to surface normal $\mathbf{g}_{\bar{2}01}$.\\
 $^d$ Fit locally and fixed during analysis due to lack of sensitivity. 
	\label{tab:domainsumparameters}
\end{table*}

The best-match model calculated functions $\varepsilon_\parallel$ and $\varepsilon_\perp$ (Eqs.~\ref{eq:uniaxialDFcomponents1},~\ref{eq:uniaxialDFcomponents2}) and inverse dielectric functions $\varepsilon^{-1}_\parallel$ and $\varepsilon^{-1}_\perp$ (Eqs.~\ref{eq:uniaxialinvDFcomponents1},~\ref{eq:uniaxialinvDFcomponents2}) are shown in Figs.~\ref{fig:DF} and~\ref{fig:invDF}, respectively. Note that imaginary parts are plotted using logarithmic scales. The results of both sum form analyses are summarized in Tab.~\ref{tab:domainsumparameters}. Note that the lowest TO mode of character $A_{\mathrm{u}}$ at approximately 155~cm$^{-1}$ was not observed due to noise, and kept fixed at its values observed for bulk crystals.

According to the six-domain effective uniaxial dielectric function model derived in Sec.~\ref{subsec:6domainmodel}, we notice that TO mode frequencies with  single crystal origin of $B_{\mathrm{u}}$ character appear in both effective dielectric functions. This observation is important because it has been determined previously by Kasic~\textit{et al.}\cite{PhysRevB.62.7365} that TO mode frequency parameters cannot be determined for electric field polarization parallel to the optic axis in optically uniaxial films with optic axis orientation parallel to the surface normal. Such cases arise, for example, in $c$-plane oriented GaN films grown on (0001) sapphire. Hence, while specific to the domain model derived here, it is fortunate that we can tie the TO frequency values for the parameters of the in-plane and out-of-plane TO summation form together. The modes with single crystal origin of $A_{\mathrm{u}}$ character only occur in the in-plane dielectric function. This is seen by observing the vertical lines indicating TO mode frequencies in  Fig.~\ref{fig:DF}. Further, we observe that the corresponding LO mode frequencies with single crystal origin of $B_{\mathrm{u}}$ character are different within the two dielectric functions (Fig.~\ref{fig:invDF}). We did not tie these frequencies together during the best-match model calculations. It was also discussed by Kasic~\textit{et al.}\cite{PhysRevB.62.7365} that LO modes can be determined with high accuracy for such optically uniaxial film configurations. Specifically, the LO modes for polarization parallel to the surface normal cause characteristic features in the ellipsometry spectra which are associated with the so-called Berreman effect.\cite{PhysRev.130.2193} This can be seen here in Fig.~\ref{fig:SE} where the film LO modes for parallel polarization (red dash-dotted lines) are generally associated with features in the spectra.  For the $(\bar{2}01)$-plane orientation of the six-fold domains, using the eigendielectric polarization model in Eq.~\ref{eq:6domainsum} and single crystal phonon mode parameters, the TO modes with $B_u-1$ and $B_u-4$ character are predicted to cause features mostly in $\varepsilon_{\perp}$, while $B_u-3$ is anticipated to appear nearly exclusively in $\varepsilon_{\parallel}$. These characteristics are confirmed when inspecting the amplitude of the imaginary parts at the respective frequencies in Fig.~\ref{fig:DF}. It is also confirmed in Fig.~\ref{fig:DF} that the LO mode frequencies of $B_{\mathrm{u}}$ character, in general, do not agree among the two inverse dielectric functions.

\begin{figure}
	\centering
\includegraphics[width=1.1\columnwidth, trim={2cm 0cm 0 0cm},clip]{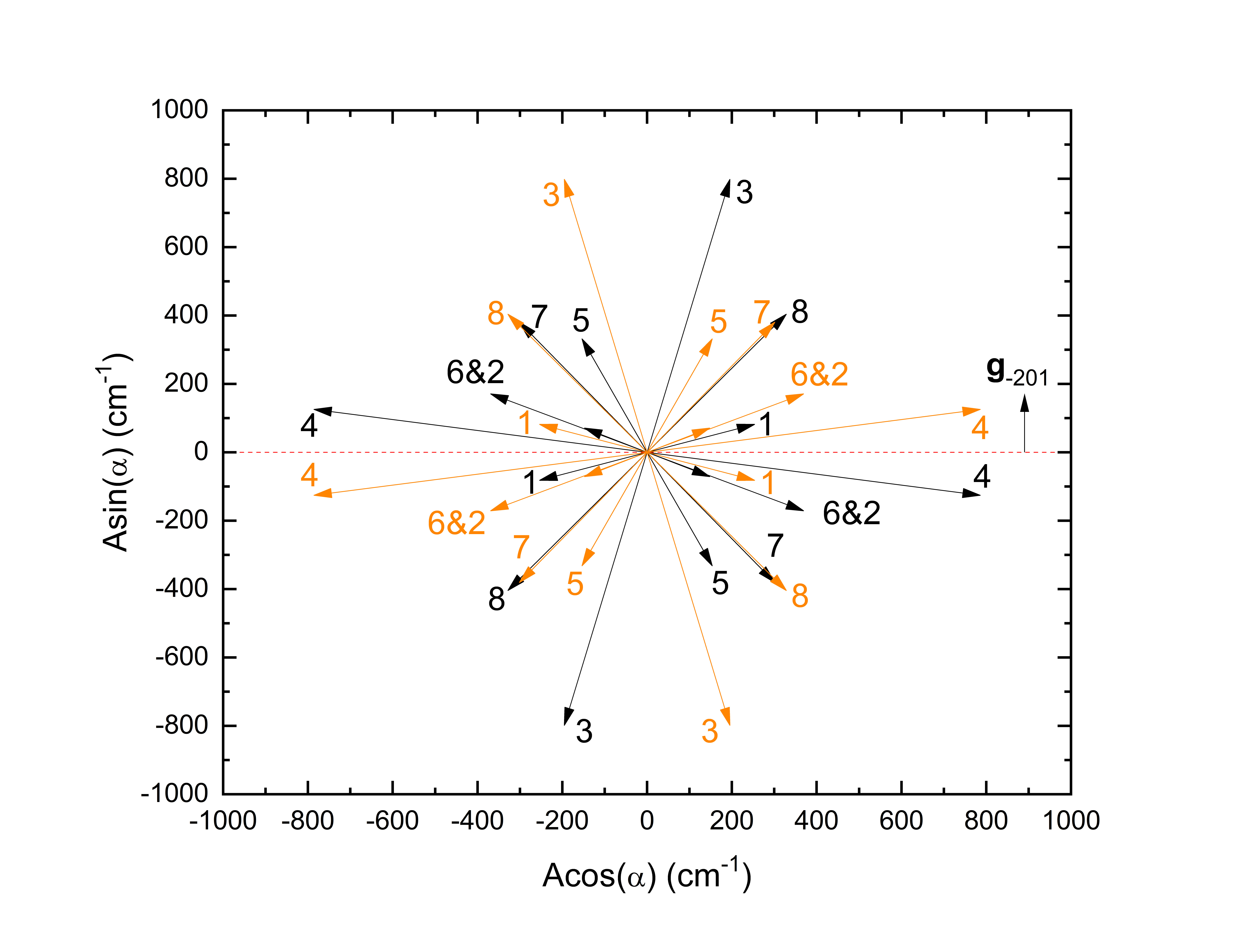}
	\caption{Schematic presentation of the single crystal phonon modes within the $\mathbf{a}-\mathbf{c}$-plane of $\beta$-Ga$_2$O$_3$ in two $(\bar{2}01)$ oriented domains, one with no (0$^{\circ}$) rotation and the other with 180$^{\circ}$ about the surface normal. Each mode is indicated by a double arrow according to its amplitude (arrow length) and dipole vector orientation (rotation in the viewing plane here). The horizontal dotted line indicates the cross section of the sample surface, to which reciprocal lattice vector $\mathbf{g}_{(\bar{2}01)}$ is perpendicular. The dipole orientations in the 0$^{\circ}$ domain (black arrows) is symmetrical about the sample surface and normal with the 180$^{\circ}$ domain (orange arrows). Hence, there is no shear element in the sum of the corresponding dielectric tensors. Each mode projects a contribution  onto the sample surface and onto the sample normal. For example, modes $B_{\mathrm{u}}-3 (l=4)$ and $B_{\mathrm{u}}-5 (l=7)$ have stronger features in the out-of-plane effective dielectric function than within the in-plane effective dielectric function. See also Fig.~\ref{fig:DF}.}
	\label{fig:orient}
\end{figure}

Figure~\ref{fig:orient} depicts a schematic of the single crystal phonon modes within the $\mathbf{a}-\mathbf{c}$-plane of $\beta$-Ga$_2$O$_3$ for the case of $(\bar{2}01)$ surface orientation. The figure displays the TO modes schematically for two oriented domains, one with no (0$^{\circ}$) rotation and the other with 180$^{\circ}$ about the surface normal. The vector $\mathbf{b}$ is collinear with the viewing direction onto the figure plane (pointing out of the page for domain zero rotation and into the plane for domain with $\pi$ rotation). A double arrow indicates the TO mode amplitude (length) and the orientation of its eigendisplacement vector. The sample surface is indicated by the horizontal dotted line and its surface normal which is the reciprocal lattice vector $\mathbf{g}_{(\bar{2}01)}$. The two domains are shown as example only for demonstration. Any other pair of oppositely oriented domains within the six domains produce the same effects. The dipole orientations in the 0$^{\circ}$ domain (black arrows) and 180$^{\circ}$ rotated domain (orange arrows) symmetrize the arrangements of TO modes across the domains. Hence, there is no shear element within the $(\hat{x}-\hat{z})$-plane in the sum of the corresponding dielectric tensors, i.e., $\hat{\varepsilon}_{xz}=\hat{\varepsilon}_{zx}=0$. This can also be verified by adding Eqs.~\ref{eq:domain0} and~\ref{eq:domain180}. Each mode then projects a contribution  onto the sample surface and onto the sample normal. For example, modes indexed $l=4$ (originating from single crystal TO mode $B_{\mathrm{u}}-3$) and $l=7$ ($B_{\mathrm{u}}-5$) possess smallest contributions of their dipoles parallel to surface and strongest perpendicular to the surface. Hence, their features are least pronounced in $\varepsilon_{\perp}$ and large in $\varepsilon_{||}$ accordingly in Fig.~\ref{fig:DF}. We note that Fig.~\ref{fig:orient} is obtained using the angular orientations of the phonon modes in the monoclinic plane presented in Ref.~\cite{Schubert2016} by adding a constant offset of 22$^\circ$ to all parameters. This offset is due to an incorrect assignment of the lattice orientation $\mathbf{a}$ relative to the sample surfaces of the samples investigated in Ref.~\cite{Schubert2016}. The arrows representing phonon displacement vectors shown in Fig.~\ref{fig:orient} now correspond to the correct orientation for $(\bar{2}01)$ domains.

Also indicated in Figs.~\ref{fig:DF} and~\ref{fig:invDF} are the frequencies of the bulk single crystal phonon modes with $A_{\mathrm{u}}$ and $B_{\mathrm{u}}$ character. Two main observations are made here. First, the TO phonon modes of the effective uniaxial film slightly shift from the bulk values, and second, the LO phonon modes are completely different both in frequency as well as in order relative to their TO counterparts.

The TO mode frequencies shift from the anticipated single crystal values. This is a surprising observation since very small strain values were observed from XRD investigations. Hence, one would assume that specifically TO phonon modes would appear close to the same frequency as for the bulk case. The latter is concluded from the fact that TO modes reflect the local strength of a lattice bond, i.e., the dynamic restoring conditions for a local molecular displacement. As seen in Fig.~\ref{fig:DF}, modes with $A_u-3$ and $B_u-1$ characters are shifted by tens of wavenumbers, and which cannot be explained by small residual amounts of strain. As discussed in our previous work,\cite{Korlacki2020} where we elucidated the effects of strain onto the lattice dynamics in $\beta$-Ga$_2$O$_3$, nonphysically large amounts of strains would be needed to warrant such large shifts. At the same time, the other modes would then experience large shifts as well, while we observe here almost no shifts for example, for modes with $A_u-1$, $B_u-6$ and $B_u-7$ character. We hypothesize here that the influence of grain boundaries may cause the observed TO mode shifts. Lattice vibrations are sensitive to small crystal dimensions due to confinement and breakdown of wavevector and polarization selection rules.\cite{Spirkoska_2008,10.1063/1.120648} A high density of grain boundaries are typically found in the ($\bar{2}01$)-plane oriented films with multiple rotation domains.\cite{Gogova2022} These irregularities and frequent interruptions of the periodic lattice may be responsible not only for the observed increased broadening of the phonon modes but may also cause a shift of their resonance frequencies. 
As a coarse estimate from XRD linewidths of the ($\bar{2}01$) peak we obtain a typical vertical grain size on order of 70~nm according to the Scherrer equation. The lateral grain size varies with distance from the substrate and was found to be tens of nanometer.\cite{Gogova2022} 

The so-called TO-LO rule dictates that TO and LO mode frequencies in cases with multiple-phonon mode branches alternate such that with ascending wavenumbers every TO mode is followed by exactly one LO mode, thus leading to a sequence TO-LO-TO-LO etc.\cite{Schubert-book} In another work we showed previously that this rule is violated for materials with monoclinic lattice structures, such as $\beta$-Ga$_2$O$_3$.\cite{Schubert2019a} There, the phonon order and reststrahlen bands of polar vibrations in crystals with monoclinic symmetry was discussed, and it was shown that $\beta$-Ga$_2$O$_3$ forms nested phonon modes for lattice displacements with $B_{\mathrm{u}}$ character, where inner and outer pairs of branched TO-LO modes lead to sequences such as TO-TO-LO-TO or TO-TO-LO-LO, etc. For example, TO mode $B_u-8$ is part of an outer mode pair followed by TO mode $B_u-7$, part of an inner mode, in ascending frequency order, followed by LO mode $B_u-7$, which is followed by TO mode $B_u-6$ and so on, hence, leading to a sequence TO-TO-LO-TO-LO-LO for the first complete set of nested modes with eigenpolarizations within the monoclinic plane. As can be seen in Fig.~\ref{fig:invDF}, where the imaginary parts peak at LO resonances, and within Tab.~\ref{tab:domainsumparameters} the order of LO modes in both dielectric functions here is of the type TO-LO-TO-LO etc. throughout. Hence, all LO modes observed here are completely different in frequency than observed within the single crystals previously. The cause for this reordering of the LO modes with respect to their corresponding TO modes is found in the long range ordering of the six-fold rotated domains. The size of the domains is much smaller than the wavelength of the associated lattice vibrations. Hence, during excitation of the effective LO modes, a nearly equal electric field expands over very large numbers of six-fold rotated domains. Within the domains, the lattice displacement causes a dielectric displacement which is non-zero at the LO modes observed here. This is simply because within the domains, the rules for the monoclinic lattice excitations must be fulfilled, and the LO modes appear at different frequencies as lined out in the previous works. Instead, the addition of the local displacements over the adjacent six-fold domains leads to a new condition when the total displacement across adjacent domains vanishes, and as a result, determines at which new frequencies the effective dielectric response reveals dielectric loss. This in itself is not a new observation, and was reported for textured crystalline films already, for example, in wurtzite-structure hexagonal boron nitride ($h$-BN) films.\cite{PhysRevB.56.13306} However, the interesting aspect here is that locally, and across grain boundaries, at the LO resonance of the film, the lattice vibration is associated with very specific, directional dielectric displacement and which is dependent on frequency, grain size, and material at the grain boundaries. For example, laterally nanostructured films with six-fold rotation domain texture of monoclinic symmetry lattice structures can offer interesting designs for nanophotonic and nanophononic device applications. Note that contrary to high-symmetry materials, such as textured wurtzite-structure $h$-BN, phonon modes in domains of low-symmetry materials have very specific directions in resonance condition, and which maybe facilitated in phononic devices for selective information transport, for example.

The broadening parameters reported in Tab.~\ref{tab:domainsumparameters} were obtained from the same best-match model calculations. Parameters for the same TO modes in both dielectric functions were coupled together, hence, no separate broadening parameters are determined for the parallel dielectric function. We observe a general increase in broadening parameter values within the TO sum forms in comparison with the TO broadening parameters reported for the bulk single crystal investigations previously.\cite{Schubert2016} This increase by up to an order of magnitude can be explained by the small grain size in the highly textured film investigated here. A comparison for the LO mode broadening parameters with data from bulk single crystal investigations is not available at this time.

\section{Conclusions}
We investigated the mid- and far-infrared optical properties of six-fold rotation domain $\bar{2}01$-oriented $\beta$-Ga$_2$O$_3$ films grown by metal-organic vapor deposition onto $c$-plane oriented sapphire substrates. We observed an effective optical uniaxial behavior with differing dielectric functions for electric field polarization parallel and perpendicular to the surface normal of the films. We find that all phonon modes known from bulk single crystal $\beta$-Ga$_2$O$_3$ occur within the spectra of the two dielectric functions of the epitaxial $\beta$-Ga$_2$O$_3$, however, the transverse optical modes are shifted and the longitudinal modes appear in new order and at noticeably different frequencies. We also observe that transverse optical modes descending from single crystal with $B_{\mathrm{u}}$ character occur in both dielectric functions, while those with $A_{\mathrm{u}}$ character only occur in the in-plane dielectric function. We analyze both functions with TO and LO sum forms and report all phonon mode model parameters for a representative film. The observed shift in TO frequencies with respect to single crystalline undoped bulk material can be explained with small grain sizes and domain wall effects, while the change in phonon mode order is caused by long range order of the six-fold domain structure where the effective LO modes appear as new conditions for vanishing dielectric displacement when integrating over large areas with multiple domains. We propose nanostructured low-symmetry films with designed texture for applications in nanophotonic and nanophononic devices.

\section*{Acknowledgments}

AM thanks Tania Paskova, Nicholas Blumenschein, and John Muth for starting me down this path by providing some of my first samples of $\beta$-Ga$_2$O$_3$ on sapphire. This work was performed within the framework of the Center for III-Nitride Technology, C3NiT-Janz\'en, supported by the Swedish Governmental Agency for Innovation Systems
(VINNOVA) under the Competence Center Program Grant No. 2022-03139. 
We further acknowledge support from the Swedish Research Council under Grants No. 2016-00889 and No. 2022-04812, the Swedish Foundation for Strategic Research under Grants No. RIF14-055 and No. EM16-0024, and the Swedish Government Strategic Research Area NanoLund and in Materials Science on Functional Materials at Link\"oping University, Faculty Grant SFO Mat LiU No. 009-00971. The work at IKZ was funded by the BMBF under Grant Nos. 03VP03712 and 16ES1084K, the European Community (Europ\"aische Fonds f\"ur regionale Entwicklung-EFRE) under Grant No. 1.8/15, and by the Deutsche Forschungsgemeinschaft under project funding Reference No. PO-2659/1-2. M.S. acknowledges support by the National Science Foundation under awards DMR 1808715, ECCS 2329940, and EPSCoR RII Track-1: Emergent Quantum Materials and Technologies (EQUATE), OIA-2044049, by Air Force Office of Scientific Research under awards FA9550-19-S-0003, FA9550-21-1-0259, and FA9550-23-1-0574 DEF, and by the University of Nebraska Foundation. A.M. and M.S. also acknowledge support from the J.~A.~Woollam Foundation.



\end{document}